\documentclass[11pt,a4paper]{article}
\usepackage{jheppub}

\usepackage{amsmath,amssymb,amsfonts, physics}
\usepackage{graphicx}
\usepackage{bm}
\usepackage[dvipsnames]{xcolor}
\usepackage[normalem]{ulem}

\usepackage{url}
\usepackage[nameinlink]{cleveref}

\usepackage{acronym}

\usepackage{orcidlink}

\usepackage{tikz}

\definecolor{tealblue}{rgb}{0.21, 0.56, 0.63}
\hypersetup{colorlinks=true,allcolors = tealblue,linktocpage=true}

\renewcommand{\figurename}{{\textbf{Figure}}}
\renewcommand{\thefigure}{{\textbf{\arabic{figure}}}}
\renewcommand{\tablename}{{\textbf{Table}}}
\renewcommand{\thetable}{{\textbf{\arabic{table}}}}

\newacro{UV}[UV]{ultraviolet}
\newacro{IR}[IR]{infrared}
\newacro{QFT}[QFT]{quantum field theory}
\acrodefplural{QFT}{quantum field theories}
\newacro{EFT}[EFT]{effective field theory}
\acrodefplural{EFT}{effective field theories}
\newacro{RG}[RG]{renormalization group}
\newacro{SM}[SM]{Standard Model}
\newacro{BSM}[BSM]{Beyond Standard Model}
\newacro{DM}[DM]{dark matter}
\newacro{ULDM}[ULDM]{ultra-light dark matter}
\newacro{QCD}[QCD]{quantum chromodynamics}
\newacro{ALP}[ALP]{axion-like particle}

\newcommand{\eg}{e.g.,}
\newcommand{\ie}{i.e.,}

\newcommand{\thcoupling}{\ensuremath{\zeta}}
\newcommand{\thcouplingFP}{\ensuremath{\zeta_\ast}}
\newcommand{\scalarmass}{\ensuremath{m_\phi}}
\newcommand{\scalarmassFP}{\ensuremath{m_{\phi,\ast}}}

\newcommand{\regulatoroperator}{\ensuremath{\mathfrak R}}

\allowdisplaybreaks

\graphicspath{{figures/}, {}}

\title{Towards theory constraints on ultralight dark matter from quantum gravity}
\author[a,b]{Gabriel Assant,\,\orcidlink{0009-0002-4471-2547}\,}
\author[a]{Astrid Eichhorn,\,\orcidlink{0000-0003-4458-1495}\,}
\author[a]{and Benjamin Knorr\,\orcidlink{0000-0001-6700-6501}\,}

\affiliation[a]{Institut f\"ur Theoretische Physik, Universit\"at Heidelberg, Philosophenweg 16, 69120 Heidelberg, Germany}
\affiliation[b]{Department of Physics and Astronomy, University of Sussex, Brighton, BN1 9QH, U.K.}

\emailAdd{eichhorn@thphys.uni-heidelberg.de}
\emailAdd{knorr@thphys.uni-heidelberg.de}
\emailAdd{g.assant@sussex.ac.uk}

\abstract{Ultralight scalar dark matter may couple to the Standard Model through dimension-five operators that contain the field-strength tensors of the gauge interactions. Recent progress in nuclear clocks is projected to increase the sensitivity to such couplings by several orders of magnitude. Future experimental constraints may even have Planck-scale sensitivity, calling for a study of such couplings in a framework that includes quantum gravity. We take a first step towards providing the theoretical constraints on such couplings that arise in asymptotically safe gravity. We find evidence that such couplings vanish in asymptotically safe gravity and are also not generated in a perturbative quantum-gravity regime that describes quantum gravity as an effective field theory.}

\begin{document}
\maketitle

\section{Introduction}\label{sec:intro}

\Ac{DM} is widely believed to be an important part of the composition of our Universe, that is needed to explain, \eg{} galaxy rotation curves, the peaks in the CMB power spectrum, some gravitational lensing observations, or structure formation, to name a few. By definition, \ac{DM} mainly interacts gravitationally, and only very weakly with the \ac{SM}. This makes it difficult to directly observe \ac{DM}. Because there has not yet been a direct detection, numerous ideas about the composition of \ac{DM} exist~\cite{Cirelli:2024ssz}.

Out of all proposals, \ac{ULDM} provides a candidate that has recently received strong attention, after extensive searches for weakly-interacting massive particles have not resulted in a detection. A strengthened experimental effort~\cite{Ferreira:2021, Hui:2021tkt, Arvanitaki:2015, Campbell:2021, Wcislo:2018, Antypas:2021, Vermeulen:2021, Filzinger:2023qqh, ADMX:2025vom, Flambaum:2022zuq}, and corresponding theoretical developments, therefore focus on the part of \ac{DM} parameter space that lies at masses much below the GeV-scale~\cite{Antypas:2022asj, Adams:2022pbo, Jaeckel:2022kwg, Dzuba:2023zcq}. A potential signature of \ac{ULDM} is that it could contribute time-dependent corrections to \ac{SM} parameters. This is because, at low enough masses, the occupation number is at least one particle per de Broglie volume, such that \ac{ULDM} can be described as a classical field, oscillating at a frequency determined by its mass. In this way, the scalar \ac{ULDM} induces oscillations of nuclear parameters like the fine-structure constant, quark masses, and the \ac{QCD} scale $\Lambda_{\text{QCD}}$~\cite{Arvanitaki:2014faa, Uzan:2010pm, Stadnik:2015kia}.

A possible low-energy interaction Lagrangian that couples \ac{ULDM} to the \ac{SM} is
\begin{align}\label{eq:interaction_Lagrangian_ULDM}
    \mathcal{L}_{\phi} = \left[ \frac{d_e}{4e^2} F_{\mu\nu} F^{\mu\nu} +\frac{d_g}{4g_s^2} G^a_{\mu\nu} G^{a\mu\nu} - d_{m_e} m_e \bar{e} e - \sum_{q=u,d} \left(d_{m_q} + \gamma_{m_q} d_g \right) m_q \bar{q} q \right] \kappa \phi \, ,
\end{align}
where the \ac{ULDM} scalar $\phi$ couples linearly to \ac{SM} fields via couplings $d_i$. These are the leading-order terms in an \ac{EFT} of \ac{ULDM} and the \ac{SM}, under the assumption that the \ac{ULDM} field is a scalar that has no internal symmetries, not even a $\mathbb{Z}_2$ reflection symmetry. Our notation deviates slightly from~\cite{Damour:2010rm, Damour:2010rp}.\footnote{We relate the strong sector in \eqref{eq:interaction_Lagrangian_ULDM} with the one of~\cite{Damour:2010rm, Damour:2010rp} by $\frac{d_g}{4g_s^2}\to-\frac{d_g \beta_s}{2 g_s}$, where $\beta_s$ is the \ac{QCD} beta function. The latter parametrization is motivated by the \ac{QCD} conformal anomaly.} The dimensionless \ac{ULDM} coupling index runs over $i=e,g,m_e,m_q$, where $\gamma_{m_q}$ are the anomalous dimensions of the up and down quarks, and $\kappa=\sqrt{4\pi}/M_{\text{Pl}}$ is the inverse reduced Planck mass. In this model, corrections to the nuclear binding energy as a function of $d_i$ have been computed, and used successfully to constrain said \ac{ULDM} couplings via composition-dependent experimental tests of the weak equivalence principle~\cite{Damour:2010rm, Damour:2010rp}.

Experimentally, oscillations of \ac{SM} parameters can also be found by comparing rates of two frequency standards which depend on \ac{ULDM} parameters differently~\cite{Damour:2010rm, Damour:2010rp, Uzan:2010pm, Stadnik:2015kia}. Unprecedented experimental opportunities are currently being opened up by the development of nuclear clocks \cite{Peik:2020cwm, Delaunay:2025lgk, Banerjee:2023bjc, Kozyryev:2018pcp}. Optical nuclear clocks keep time using nuclear transitions. 
Recently, the Thorium $^{229}\text{Th}$ isotope was identified as a preferred candidate for such experiments~\cite{Kraemer03042025, Kazakov:2012cao, Kraemer:2022gpi, Caputo:2024doz, Thirolf:2019ocm, Higgins:2024fir, Flambaum:2008ij, Berengut:2009zz, Beeks:2022dnl, Elwell:2024qyh, Tiedau:2024obk, Thirolf:2024xlx, Flambaum:2025xqs, Zhang:2024ngu, Flambaum:2023bnw}. Indeed, the cancellation of electromagnetic and strong contributions to the Thorium nuclear binding energy leads to a low-lying isomeric transition~\cite{Tkalya:2000yb, Flambaum:2006ak}, making it the only transition accessible by state-of-the-art vacuum ultraviolet lasers. Excited states of Thorium nuclei have already been achieved with different host crystals~\cite{Tiedau:2024obk, Elwell:2024qyh}, paving the way towards Thorium-based laser spectroscopy and optical nuclear clocks. This corresponds to an improvement of three orders of magnitude compared to the indirect measurement of the isomeric transition energy via the detection of the radiative decay~\cite{Kraemer:2022gpi}. Recent experiments also managed laser excitations of Thorium using single modes of a vacuum-ultraviolet frequency comb~\cite{Zhang:2024ngu}. 

On the theoretical side, the low-lying nature of the Thorium nuclear transition improves the sensitivity to probes of quantum electrodynamics and \ac{QCD}. The latter is improved by eight orders of magnitude compared to previous experiments~\cite{Flambaum:2008ij, Berengut:2009zz}. In other words, oscillations in $\Lambda_{\text{QCD}}$ due to \ac{ULDM} imply a larger modification of the Thorium nuclear transition frequency, relative to the modifications induced in the nuclear transitions of other elements. Sensitivites to scalar and pseudoscalar \ac{ULDM} couplings may even reach the (inverse) Planck scale \cite{Fuchs:2024xvc, Kim:2022ype}.

Any experiment that reaches Planck-scale sensitivity is also an opportunity to tackle one of the most fiendish challenges in fundamental physics, namely the confrontation of quantum gravity theory with experimental data. Because gravity couples to all forms of energy and matter, there is a general expectation that quantum gravity fluctuations induce matter interactions upon being integrated out. This can be thought of in complete analogy to integrating out any other field in the \ac{UV} and thereby generating higher-order interactions in the \ac{EFT}. Thus, one would generically expect that the couplings $d_e, d_g$ etc.~in Eq.~\eqref{eq:interaction_Lagrangian_ULDM} are $\mathcal{O}(1)$, given that the scale in $\kappa$ is already the Planck scale.

Different proposals for quantum gravity theories differ according to which quantum fluctuations they contain. For instance, the most conservative approach to quantum gravity, asymptotically safe gravity, only contains quantum fluctuations of the metric, which one can, loosely speaking, think of as virtual gravitons. By contrast, string theory contains additional fields in its \ac{EFT} description, and, at the fundamental level, features quantum fluctuations of a non-local object, the string.

However, the case for a physical difference between such distinct theoretical proposals is not as clear-cut as one might imagine, even though strong arguments for physical differences can be made \cite{Basile:2025zjc}. It is, however, not inconceivable that the differences are at the level of the mathematical formulation, but not at the level of its physical implications.\footnote{In the framework of the so-called swampland~\cite{Vafa:2005ui, Ooguri:2006in}, this would correspond to a universal swampland~\cite{Eichhorn:2024rkc}. Completely distinct relative swamplands have, however, also been argued for \cite{Basile:2025zjc}.} Insight into such physical implications is therefore urgently needed, making the development of quantum-gravity phenomenology a pressing issue and the connection of quantum gravity to experiment mandatory~\cite{Addazi:2021xuf}. 

Here, we contribute to this effort by calculating for the first time whether couplings of the type $d_e$ and $d_g$ are present in asymptotically safe gravity. In this setting, quantum gravity can be formulated as a \ac{QFT}, without the need to introduce extra dimensions or superpartners. Instead, the theory is based on a fundamental symmetry, namely scale symmetry, which arises due to quantum fluctuations and is realized at trans-Planckian scales.

At the technical level, this proposal retains the spacetime metric as the fundamental mediator of gravitational interactions. Scale symmetry is achieved at an interacting \ac{UV} fixed point of the \ac{RG} flow. Since the seminal work~\cite{Reuter:1996cp}, the existence of the \ac{UV} fixed point has been established beyond reasonable doubt in a wide range of approximations and systems. The pure gravity Euclidean fixed point is reviewed in~\cite{Reuter:2019byg, Pereira:2019dbn, Eichhorn:2020mte, Bonanno:2020bil, Pawlowski:2020qer, Saueressig:2023irs, Knorr:2022dsx, Pawlowski:2023gym, Basile:2024oms}. The (physically more relevant) Lorentzian regime of gravity is subject to more recent studies (and related formal developments~\cite{DAngelo:2023tis}), which indicate that Euclidean results may carry over to the Lorentzian regime~\cite{Bonanno:2021squ, Fehre:2021eob, Saueressig:2023tfy, DAngelo:2023wje, Ferrero:2024rvi, Saueressig:2025ypi, DAngelo:2025yoy, Pawlowski:2025etp}.

At the \ac{UV} fixed point, much like in statistical physics, the classical power-counting scaling of operators in the fundamental Lagrangian is corrected by anomalous dimensions. The resulting quantum-mechanical operator scaling informs us if the corresponding coupling is a free parameter of the underlying theory, or is predicted instead. In this way, couplings of \ac{DM} to \ac{SM} matter have already been constrained \cite{Eichhorn:2017als, Eichhorn:2020kca, Hamada:2020vnf, Eichhorn:2020sbo, Reichert:2019car, deBrito:2021akp, Eichhorn:2021tsx, deBrito:2023ydd, Kowalska:2020zve}.\footnote{Asymptotic safety may also exist in (dark) matter models without gravity, and constrain the couplings in such models~\cite{Sannino:2014lxa, Eichhorn:2018vah, Cai:2019mtu}.}

Inspired by \eqref{eq:interaction_Lagrangian_ULDM}, we consider the interaction Lagrangian 
\begin{align}\label{eq:our_interaction_Lagrangian}
      \mathcal{L}_{\phi} = \frac{1}{4} k^{-1} \, \thcoupling \, \phi \, F^{\mu\nu} F_{\mu\nu} \, .
\end{align}
Here, $k$ is the coarse-graining \ac{RG} scale, which is a momentum scale, so that the coupling \thcoupling{} is dimensionless. Our calculation of the gravitational contribution to this coupling is insensitive to whether the field strength is Abelian or non-Abelian, and thus holds for the \ac{QCD} and the QED term. This is intuitively understood since gravity couples universally and is blind to internal symmetries. We give a more technical argument when explicitly computing the \ac{RG} running of the dimension-five coupling in \cref{sec:approx}. We will show that the coupling $\thcoupling$ is predicted to vanish in asymptotically safe gravity leading to the prediction $d_e=0$ and $d_g=0$.

This paper is structured as follows: We introduce asymptotic safety and its predictive power, review the state of the art and highlight three important aspects (global symmetries, near-perturbativity and the fate of dimension-five operators) in \cref{sec:ASgravmatsystems}. We also introduce the functional \ac{RG} and present our setup in that section. In \cref{sec:results}, we present our results, both for asymptotic safety as well as the \ac{EFT} of gravity, and in \cref{sec:conclusions}, we present conclusions and an outlook.

\section{Asymptotically safe gravity-matter system}\label{sec:ASgravmatsystems}

Asymptotic safety relies on a symmetry at (trans-)Planckian scales, namely scale symmetry. This symmetry is inherently quantum, in that it arises due to the contributions of quantum fluctuations to the effective interactions of the theory. It implies that all dimensionless combinations of couplings must be constant.\footnote{Note that this is a different requirement from setting dimensionful couplings to zero, which one would impose in a classically scale-invariant theory.} At the technical level, this is encoded in an \ac{RG} fixed point. Unlike the asymptotically free \ac{RG} fixed point of \ac{QCD}, an \ac{RG} fixed point in gravity-matter systems is necessarily interacting, \ie{} at least some couplings are nonzero.

In this section, we first review the mechanism of how asymptotic safety solves the issue of predictivity that is caused by perturbative non-renormalizability. We then review the state of the art in the field, and discuss several relevant aspects that explain our choice of approximation, as well as our focus on Abelian gauge fields. Finally, we spell out the concrete approximations that we made to arrive at our results.

\subsection{Predictivity in asymptotically safe gravity-matter systems}\label{sec:predictivity}

Gravity is famously perturbatively non-renormalizable \cite{tHooft:1974toh}. This is both true for pure gravity, as shown by the two-loop divergence \cite{Goroff:1985sz, Goroff:1985th, VANDEVEN}, and gravity coupled to matter \cite{tHooft:1974toh}. This implies a loss of predictivity beyond the Planck scale, as at every loop order, new counterterms have to be introduced to absorb divergences, which are not of the form of the original action. As a consequence, infinitely many free parameters have to be introduced.\footnote{We emphasize that this is only an issue when discussing gravity as a fundamental theory -- the gravitational \ac{EFT} works extremely well at scales below the Planck scale \cite{Bjerrum-Bohr:2002gqz}.} This can be remedied via a \ac{UV} fixed point in the \ac{RG} running of gravitational couplings. The requirement of all couplings originating from a fixed point relates the values of couplings to each other, restoring predictivity even for a setting with infinitely many higher-order terms. For this, the beta functions of all couplings $g_i$,
\begin{equation}
    \beta_{g_i} \equiv k \partial_k g_i \, ,
\end{equation}
where $k$ is the \ac{RG} scale introduced above, have to vanish,
\begin{equation}
    \text{fixed point:} \qquad \beta_{{g}_i}\big\lvert_{{g}_{i,\ast}} = 0 \, .
\end{equation}
Here ${g}_{i,\ast}$ denotes the fixed-point values of couplings. The index $i$ runs over all types of interactions present in the theory, with couplings ${g}_i$. The fixed-point condition involves dimensionless couplings. Hence, we always have to define dimensionless couplings by rescaling any dimensionful couplings with appropriate powers of the \ac{RG} scale $k$.

While it is immediately obvious that a requirement such as $\beta_{g_i}=0$ is likely to fix the values of couplings at the fixed point, it is less obvious that relations between couplings persist, once the \ac{RG} flow has left the fixed-point regime. We will now explain the mechanism that generates predictivity in a theory for which the \ac{UV} is determined by an interacting fixed point, even if the \ac{IR} does not necessarily correspond to a scale-symmetric theory.

At interacting fixed points where couplings take non-vanishing values $g_{i,\ast}\neq0$, quantum fluctuations change the scaling dimension of the interactions. This anomalous scaling is encoded in the stability matrix $M$,
\begin{equation}
    M_{ij}=\frac{\partial \beta_{g_i}}{\partial g_j}\bigg\lvert_{g_\ast} \, , \qquad \text{eigv}(M)=-\theta_i \, .
\end{equation}
The eigenvalues of this matrix are the universal critical exponents $\theta_i$, and they govern the scaling of couplings near the fixed point. This can be seen by solving the linearized beta functions about a fixed point. To illustrate this, for a single coupling $g$ about a fixed point $g_\ast$, we have
\begin{equation}
    g = g_\ast + c \, \left( \frac{k}{k_0} \right)^{-\theta} + \dots \, ,
\end{equation}
where $k_0$ is an arbitrary reference scale, and $c$ is an integration constant. From this, we can see that the real part of the critical exponent determines if an operator is relevant or irrelevant at a given fixed point. If $\theta_i>0$, the operator is relevant. Under the \ac{RG} flow towards the \ac{IR} (that is $k\to0$) the system may depart from scale symmetry along the corresponding direction, i.e., the distance of this coupling to its fixed-point value changes. Thus, the value of the corresponding coupling in the \ac{IR} is a free parameter of the theory (corresponding to the integration constant $c$ above) and must be fixed by experiments. Conversely, if $\theta_i<0$, the operator is irrelevant and the value of the corresponding coupling is predicted in the \ac{IR}, as a function of the relevant couplings. This follows, because the constant of integration, $c$, associated to a coupling with $\theta<0$ does not enter the low-energy values of couplings, i.e., it is an \emph{irrelevant} parameter. Hence, predictivity in quantum gravity is restored by a fixed point of the \ac{RG} flow with a \emph{finite} number of relevant directions. 

There is by now convincing evidence that a fixed point with no more than three relevant directions in the gravitational coupling space (more if matter is added, dependent on the type of matter interactions) exists, as found in numerous studies in different approximations of the Euclidean gravitational \ac{RG} flow~\cite{Lauscher:2002sq, Codello:2008vh, Benedetti:2009rx, Benedetti:2009iq, Groh:2011vn, Rechenberger:2012pm, Falls:2013bv, Falls:2014tra, Gies:2016con, Falls:2017lst, Falls:2018ylp, Falls:2020qhj, Kluth:2020bdv, Knorr:2021slg, Baldazzi:2023pep}.

We will study the impact of trans-Planckian quantum gravity fluctuations on the critical exponents of the matter couplings appearing in \eqref{eq:our_interaction_Lagrangian}. From this we will predict the value of the low-energy experimentally constrained couplings of \eqref{eq:interaction_Lagrangian_ULDM}.

We will continue by reviewing how asymptotic safety constrains the matter sector that can be consistently coupled to gravity in the sense of admitting a \ac{UV} completion. Then, we focus on several aspects of gravity-matter systems that play an important role for our study. First, global, internal symmetries of matter systems are -- according to the current state of the art in the field -- preserved under the impact of quantum gravity fluctuations. This constrains the general structure of beta functions, also in our study. Second, asymptotically safe gravity-matter systems, defined at a partially interacting fixed point, exhibit a scaling spectrum that is close to Gaussian, or ``near-perturbative''. Third, as a consequence of the first two aspects, dimension-five operators typically feature vanishing fixed-point values and do not correspond to relevant couplings. Therefore, the resulting phenomenology is shaped by vanishing dimension-five couplings.

\subsection{State of the art in asymptotically safe gravity-matter systems}\label{sec:ASstateofart}

Based on the extensive evidence provided for asymptotic safety in pure gravity (see \cref{sec:intro}), we will consider it as a given that quantum gravity on its own is asymptotically safe. The situation with matter is less clear, in particular in the limit of large number of matter fields, but there is good evidence that asymptotic safety persists, if the number of matter fields is not too large \cite{Dona:2013qba, Meibohm:2015twa, Biemans:2017zca, Alkofer:2018fxj, Wetterich:2019zdo}. This appears to include the case of the \ac{SM} \cite{Dona:2013qba, Meibohm:2015twa, Biemans:2017zca, Alkofer:2018fxj, Wetterich:2019zdo, Pastor-Gutierrez:2022nki}, see also \cite{Eichhorn:2022gku} for a comprehensive overview of the literature. The status of asymptotically safe gravity-matter systems is reviewed in \cite{Eichhorn:2018yfc, Eichhorn:2022jqj, Eichhorn:2022gku, Eichhorn:2023xee}. Here we provide a brief summary that focuses on the impact of quantum fluctuations of gravity on matter.

Quantum-gravity fluctuations preserve asymptotic freedom for the non-Abelian gauge interactions \cite{Daum:2009dn, Folkerts:2011jz, Christiansen:2017cxa} and solve the Landau-pole problem in the Abelian hypercharge sector of the \ac{SM}~\cite{Harst:2011zx, Christiansen:2017gtg, Eichhorn:2019yzm, DeBrito:2019gdd, deBrito:2022vbr}. This results in an upper bound on the Abelian gauge coupling~\cite{Eichhorn:2017lry}, which constitutes a testable prediction from asymptotic safety, albeit one that is currently subject to large systematic uncertainties. Additionally, constraints on photonic self-interactions  and the non-minimal coupling between photons and gravity have been derived \cite{Christiansen:2017gtg, Eichhorn:2021qet, Knorr:2024yiu}.

The ratio of Higgs mass to the electroweak scale, which is a free parameter in the \ac{SM}, is fixed in asymptotic safety~\cite{Shaposhnikov:2009pv, Pawlowski:2018ixd, Eichhorn:2021tsx}, and may, within the systematic uncertainty of the theoretical prediction as well as the experimental uncertainty on the top quark mass, be compatible with the experimental results~\cite{Pastor-Gutierrez:2022nki}.

Yukawa couplings are also affected by quantum fluctuations of gravity~\cite{Eichhorn:2016esv, Eichhorn:2017eht, deBrito:2025nog} and bounded from above. As a consequence, the masses of the top and bottom quark may be constrained in asymptotic safety~\cite{Eichhorn:2017ylw, Eichhorn:2018whv, Eichhorn:2025sux}. The quantum-gravity effects on Yukawa couplings also give rise to a mechanism that keeps neutrinos dynamically light by suppressing their Yukawa coupling to the Higgs field~\cite{Held:2019vmi, Kowalska:2022ypk, Eichhorn:2022vgp, Chikkaballi:2023cce, deBrito:2025ges}.

Three generations of quarks including their mixing were first considered in~\cite{Alkofer:2020vtb}, finding first indications that a \ac{UV} completion may also proceed through a fixed-point cascade. More recently, it was found that, in a parameterized treatment of gravitational fluctuations, such a fixed-point cascade may explain observed structures in the mixing matrices for quarks and leptons~\cite{Eichhorn:2025sux} and predict a near-diagonal quark mixing matrix.

\Ac{BSM} settings have also been explored, and, in general, strong predictive power of asymptotic safety has been found. As a consequence, some \ac{BSM} settings are ruled out, and others have a reduced set of free parameters compared to the same models in an \ac{EFT} setting, see~\cite{Eichhorn:2017als, Reichert:2019car, Eichhorn:2020kca, Kowalska:2020zve, Eichhorn:2020sbo, deBrito:2023ydd, Chikkaballi:2025pnw} for examples with \ac{DM} and \cite{Eichhorn:2017muy,Eichhorn:2019dhg} for other \ac{BSM} settings, such as Grand Unified Theories.

All the above results are obtained in Euclidean quantum gravity. The question of Lorentzian signature in asymptotically safe gravity matter systems is under investigation, with first results being highly encouraging with regards to the existence and properties of an asymptotically safe fixed point~\cite{Korver:2024sam, Pastor-Gutierrez:2024sbt, Kher:2025rve}.

\subsection{Global symmetries in asymptotically safe gravity}\label{sec:global_symmetries}

In particle-physics phenomenology, global symmetries play a central role. However, there is a conjecture that global symmetries cannot be preserved by quantum gravity. This conjecture has its origin in black-hole physics and has been found to hold in string theory \cite{Banks:1988yz, Giddings:1987cg, Lee:1988ge, Abbott:1989jw, Coleman:1989zu, Kamionkowski:1992mf, Holman:1992us, Kallosh:1995hi, Banks:2010zn}. Nowadays, it underlies many of the string-inspired swampland conjectures \cite{Vafa:2005ui, Ooguri:2006in, Brennan:2017rbf, Palti:2019pca, vanBeest:2021lhn, Grana:2021zvf, Agmon:2022thq}. The conjecture implies that global symmetries either have to be gauged, or that they are violated by interactions with a Planck-scale suppression. Nevertheless, there are examples of quantum gravity theories, in which the conjecture does not hold \cite{Harlow:2020bee}. 

In the case of asymptotic safety, the situation is not ultimately clear, see \cite{Eichhorn:2022gku} for an overview. All results to date show that global symmetries are preserved \cite{Eichhorn:2011pc, Eichhorn:2012va, Labus:2015ska, Percacci:2015wwa, Eichhorn:2017eht, Eichhorn:2020sbo, Ali:2020znq, deBrito:2021pyi, Eichhorn:2021qet, Laporte:2021kyp}, as quantum gravity fluctuations are integrated out, even including global spacetime symmetries such as CPT \cite{Eichhorn:2025ilu}.\footnote{The results in these papers are not necessarily cast in the light of the no-global-symmetries conjecture. However, they all show that global symmetries of the kinetic terms for matter fields are respected, when quantum gravity induces new interactions.} This includes examples for scalar, fermionic as well as vector matter, and various internal global symmetry groups. However, these results are subject to systematic approximations (truncations of the dynamics to finite order in an \ac{EFT}-like expansion) and are performed in Euclidean signature. This aspect may be what prevents black-hole configurations in the path integral to have their full, symmetry-violating effect on matter. However, it may also be that the asymptotically safe dynamics, which is not Einstein Hilbert, produces a different black-hole thermodynamics \cite{Basile:2025zjc} and results in the preservation of global symmetries through dynamical suppression of black-hole configurations in the path integral \cite{Borissova:2020knn, Borissova:2024hkc}.

In practice, in calculations in the Euclidean regime, one can rely on the preservation of global symmetries. Thus, for the purposes of this work, we can anticipate that, because of the preservation of global shift-symmetry for the scalar field, the coupling $\thcoupling$ vanishes at the asymptotically safe gravity-matter fixed point.\footnote{It is also compatible with the preservation of global symmetries that there is a second fixed point, at which the coupling is finite. However, the existence of a fixed point at vanishing coupling (or, more precisely, the absence of fluctuations that drive the \ac{RG} flow out of the space of interactions that preserve shift symmetry), is guaranteed.} We will, however, not use this as an \emph{input} of our work. Rather, it will be a result that lends further support to the violation of the no-global-symmetries conjecture in asymptotic safety.

\subsection{Near-perturbativity of asymptotic safety}\label{sec:near_pert}

The \ac{SM} is perturbative up to the Planck scale. There is by now compelling evidence that this property survives (with some qualifications, see below) beyond $M_{\text{Pl}}$, once the \ac{SM} is augmented by asymptotically safe quantum gravity \cite{Eichhorn:2017ylw, Eichhorn:2017lry, Eichhorn:2018whv, Alkofer:2020vtb, Pastor-Gutierrez:2022nki, Kowalska:2022ypk, Pastor-Gutierrez:2024sbt, Eichhorn:2025sux}. Perturbativity does not necessarily refer to small coupling values, because couplings can be rescaled arbitrarily in a Lagrangian. Instead, a measure of near-perturbativity at a \ac{UV} fixed point is given by the amount by which scaling exponents deviate from their canonical values, which they have at a fully perturbative, free fixed point. 

Indeed, for many gravity-matter systems, scaling exponents remain close to canonical ones, for various gravity-matter systems \cite{Eichhorn:2018akn,Eichhorn:2020sbo}, and symmetry-identities for the gauge symmetry in gravity indicate a near-perturbative behavior \cite{Eichhorn:2018akn, Eichhorn:2018ydy, Eichhorn:2018nda}. Near-perturbative behavior also holds true for theories of pure gravity, endowed with higher-curvature invariants \cite{Falls:2013bv, Falls:2014tra, Falls:2017lst, Falls:2018ylp, Kluth:2020bdv, Baldazzi:2023pep, Becker:2024tuw}.

Based on the notion of near-perturbativity, i.e., that critical exponents closely follow the canonical scaling of operators, $\theta_i\approx d_{\tilde{g}_i}$, we devise truncation schemes that systematically account for higher-order operators in gravity-matter systems. Based on this, computing the running of dimension-five operators, such as the \ac{ULDM}-photon operator in \cite{Fuchs:2024xvc}, is the natural next step in gravity-matter studies \cite{deBrito:2021akp, deBrito:2025ges}. In addition, dimension-five operators are the most interesting classically irrelevant operators in the infinite tower of operators compatible with symmetries. This is because their canonical mass dimension is closest to the boundary $\theta=0$, where couplings are marginal, so quantum fluctuations are more likely to make them relevant at \ac{RG} fixed points than higher-order operators.

\subsection{Dimension-five and higher operators}\label{sec:dim_five}

In \ac{BSM} settings, dimension-five operators appear naturally and some of them have been studied, including a coupling of \acp{ALP} to the electromagnetic field strength \cite{deBrito:2021akp}, as well as the Weinberg operator for neutrinos \cite{deBrito:2025ges}. Both operators turn out to be irrelevant at gravity-matter fixed points, but the gravitational parameter space contains a boundary beyond which they could become relevant. This boundary characterizes a more strongly-coupled, non-perturbative quantum-gravity regime and lies relatively far from the best estimates for fixed-point values.

The existing results raise the question whether there is a more general statement that can be made about dimension-five operators in asymptotic safety. Both known examples satisfy that the shift in the critical exponent can be \emph{towards relevance}, but not large enough to render the corresponding coupling relevant.\footnote{For the Weinberg operator, there are also smaller regions of the parameter spaces where the shift is towards irrelevance.}

In this work, the coupling $\thcoupling$ is both of phenomenological as well as more structural interest, because it serves as another test-case of the conjecture that the properties of dimension-five operators described above may generalize.

\subsection{Renormalization Group}\label{sec:RG}

\subsubsection{Flow equation}\label{sec:flow_equation}

The modern understanding of the \ac{RG} is based on Wilson's idea of integrating out momentum modes shell by shell. One specific implementation of this is the functional \ac{RG}, which is formulated in terms of an effective action, and where modes in the path integral below a fiducial momentum scale $k$ are suppressed by a regulator. When $k\to\infty$, no modes are integrated out, and the action is the microscopic one. Upon lowering this scale $k$ to zero, all fluctuations are integrated out, and one arrives at the standard effective action. This scale-dependent effective action, $\Gamma_k$, fulfills a formally exact \ac{RG} equation~\cite{Wetterich:1992yh, Morris:1993qb, Ellwanger:1993mw},
\begin{equation}\label{eq:wetterich_eq}
    k \partial_k \Gamma_k = \frac{1}{2} \text{Tr} \left[ \left( \Gamma_k^{(2)} + \regulatoroperator_k \right)^{-1} \, k \partial_k \regulatoroperator_k \right] \, .
\end{equation}
In this, $\regulatoroperator_k$ is the above-mentioned regulator, $\Gamma_k^{(2)}$ is the second functional derivative of $\Gamma_k$, and the functional trace includes a sum over all fields as well as over eigenvalues of the operator within it. In our system, the trace will sum over the fields $\phi,A_\mu,g_{\mu\nu}$, and the sum over eigenvalues can be written as an integral over the loop momentum.

In practice, the exact equation \eqref{eq:wetterich_eq} has to be approximated, as in general the \ac{RG} flow generates all terms compatible with the underlying symmetries of the theory. In the following subsection, we will explain our approximation for $\Gamma_k$.

We highlight that asymptotic safety shares challenges with other quantum-gravity approaches \cite{deBoer:2022zka, Buoninfante:2024yth} and that, despite it being mostly associated with functional \ac{RG} techniques, there is also significant progress using lattice techniques in causal and Euclidean dynamical triangulations, see~\cite{Loll:2019rdj, Loll:2022ibq, Ambjorn:2024pyv, Schiffer:2025cqc} for reviews.

\subsubsection{Approximations}\label{sec:approx}

We will now explain our ansatz for solving \eqref{eq:wetterich_eq}. For this, we split $\Gamma_k$ into a gravitational and a matter contribution. In the gravitational sector, we approximate the dynamics by the Einstein-Hilbert action,
\begin{equation}\label{eq:EH}
    \Gamma_k^\text{grav} = \frac{1}{16\pi k^{-2} g} \int \text{d}^4x \, \sqrt{\det g} \, \left( 2k^2\lambda - R \right) \, .
\end{equation}
Within the functional \ac{RG}, all couplings, like the Newton's constant $g$ and the cosmological constant $\lambda$, become $k$-dependent, and are rescaled by appropriate powers of $k$ to be dimensionless.

The action \eqref{eq:EH} has to be supplemented by a gauge fixing. To do so, we use the background field method and split the metric $g_{\mu\nu}$ into a fixed but arbitrary background $\bar g_{\mu\nu}$\footnote{Eventually, we will choose a flat background, as this will be sufficient to extract the beta functions that we are interested in.} and fluctuations $h_{\mu\nu}$ via
\begin{equation}
    g_{\mu\nu} = \bar g_{\mu\nu} + \sqrt{32\pi \, k^{-2}\, g \, Z_h} \, h_{\mu\nu} \, .
\end{equation}
This normalization gives the standard bosonic mass dimension to the metric fluctuation, $[h_{\mu\nu}]=1$. We have also introduced the graviton wavefunction renormalization $Z_h$, which defines the anomalous dimensions via
\begin{align}
    \eta_h = -k\,\partial_k \ln(Z_h) \, .
\end{align}
With this, we can define a gauge fixing condition for the fluctuation. We use a one-parameter family of linear covariant gauges,
\begin{equation}
    \mathcal F_\mu = \left( \delta_\mu^{\phantom{\mu}\alpha} \bar D^\beta - \frac{1 + \beta}{4} \bar g^{\alpha\beta} \bar D_\mu \right) h_{\alpha\beta} \, .
\end{equation}
The gauge fixing condition is implemented via the standard Faddeev-Popov method. We will use the gauge parameter $\beta$ to analyze the stability of our system. Physical results should of course not depend on the choice of gauge, if no approximation is made in the calculation. The approximation introduced by the choice of truncation results in a gauge-dependence even in physical results (e.g., critical exponents). This dependence is expected to decrease as the quality of the truncation increases and all dynamically important interactions are accounted for. Thus, an estimate for systematic uncertainties can be extracted from the gauge dependence of physical results.

We take the Landau limit to implement the gauge fixing condition sharply.\footnote{This has the additional benefit that any potential $k$-dependence of the gauge parameters $\alpha$ and $\beta$ is absent~\cite{Litim:1998qi, Knorr:2017fus}.} Since we do not compute the beta functions of the gravitational couplings in this work, the gravitational Faddeev-Popov ghosts do not play a role as they do not couple to matter.

In the matter sector, we consider an uncharged massive scalar $\phi$ to represent the \ac{ULDM} field, and couple it to a photon via a dimension-five operator. We will approximate this sector in a minimal way, relying on near-perturbativity of asymptotically safe gravity discussed in \autoref{sec:near_pert} for this approximation to be justified. We also rely on the fact that near-perturbative quantum fluctuations in the Euclidean regime do not break any global symmetries of the matter sector. As a consequence, we will focus on the scaling dimensions of the coupling $\thcoupling$ and the mass $\scalarmass$ of the scalar field. These do not receive contributions from higher-order terms in the scalar potential (except from the $\phi^4$-term, which for simplicity we neglect), nor from non-minimal couplings, at the fixed point at which all these couplings vanish in order to preserve the shift symmetry and $\mathbb{Z}_2$ symmetry of the scalar field.

Phenomenologically, the dimension-five operator that couples the scalar to the non-Abelian field strength of the SU(3) gauge interactions is more interesting, compared to the coupling to the Abelian gauge field. In asymptotically safe gravity, the gravitational contributions to the scaling dimension of both dimension-five operators are actually identical. This relies on the coupling structure of quantum gravity, which couples to any form of energy or mass, but is ``blind'' to internal symmetries.\footnote{Additional contributions from quantum gravity and gauge-field fluctuations may exist at finite values of the respective gauge couplings. However, for non-Abelian gauge interactions it is well-established that they vanish at the asymptotically safe fixed point \cite{Daum:2009dn, Folkerts:2011jz, Christiansen:2017cxa, deBrito:2022vbr}, whereas for Abelian interactions there are two possible fixed points \cite{Harst:2011zx, Eichhorn:2017lry}, one of which lies at a vanishing gauge coupling. Our work analyzes the properties of this latter fixed point.}

\begin{figure}[t]
\centering
\includegraphics[width=.7\linewidth]{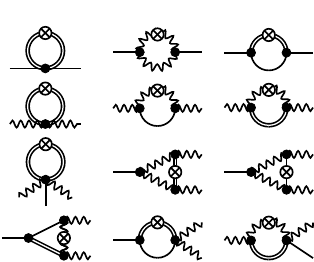} 
\caption{The first line contributes to the running scalar field mass \scalarmass{} and anomalous dimension $\eta_\phi$. The second line contributes to the gauge anomalous dimension $\eta_A$. Finally, the third and fourth lines give the running of \thcoupling{}. Double lines represent graviton fields, single lines scalar fields, and wiggly lines gauge fields. All vertices and propagators are fully dressed. The cross represents the regulator insertion $k\partial_k \regulatoroperator_k$, which must be applied to all loop propagators in turn. Symmetrization with respect to exchange in external momenta is understood.}
\label{fig:Full_Flows_matter}
\end{figure}

The preceding discussion entails the approximation
\begin{equation}\label{eq:matter_ansatz}
\begin{aligned}
    \Gamma_k^\text{mat} = \int \text{d}^4x \, \sqrt{\det g} \, \Bigg[ \frac{Z_\phi}{2} (\partial^\mu\phi)(\partial_\mu \phi)
    &+ \frac{Z_\phi}{2} k^2\,\scalarmass^2 \, \phi^2 \\
     &+ \frac{Z_A}{4} F^{\mu\nu} F_{\mu\nu} + \frac{Z_A \sqrt{Z_\phi}}{4\, k}  
    \thcoupling \, \phi \, F^{\mu\nu} F_{\mu\nu} \Bigg] \, .
\end{aligned}
\end{equation}
Here, we already introduced factors of the corresponding wave-function renormalizations $Z_{\phi,A}$.  To \eqref{eq:matter_ansatz}, we have to add a gauge fixing for the photon, and we pick the Lorenz gauge $D^\mu A_\mu=0$ in the Landau limit. The Abelian Faddeev-Popov ghosts do not contribute to any 
beta functions that we compute, and are neglected. We also introduce anomalous dimensions in the matter sector via
\begin{equation}
\eta_{\phi,A} =-k\, \partial_k \ln \left(Z_{\phi,A}\right)\,.
\end{equation}
Summarizing, we will compute fixed points for the couplings $\scalarmass$ and $\thcoupling$ while treating $g$ and $\lambda$ as parameters. Moreover, we will also compute $\eta_A$ and $\eta_\phi$, but set $\eta_h=0$. To compute the beta functions and anomalous dimensions, we take functional derivatives of the flow equation \eqref{eq:wetterich_eq}. A diagrammatic representation of the corresponding diagrams -- which resemble Feynman diagrams, but are \ac{UV} and \ac{IR} finite due to the insertion of the regulator -- is given in \autoref{fig:Full_Flows_matter}. The beta functions were computed using the \emph{Mathematica} package \emph{xAct} \cite{Martin-Garcia:2007bqa,Brizuela:2008ra,Martin-Garcia:2008yei,Martin-Garcia:2008ysv,Nutma:2013zea}.

The final missing ingredient for the computation is to specify the regulator. We adapt the regulator to the respective two-point function,
\begin{equation}\label{eq:spectrally_adjusted_reg}
    \regulatoroperator_k(p^2) = \left( \Gamma_k^{(2)}(p^2) - \Gamma_k^{(2)}(0) \right)_{h, \phi, A = 0} \, r_k(p^2 / k^2) \, .
\end{equation}
%
Explicitly, this corresponds to
\begin{equation}
\begin{aligned}
    \regulatoroperator_k^{h,\,\mu\nu\rho\sigma}(p^2) &= p^2 \left[ \frac{1}{2} \left( \bar g^{\mu\rho} \bar g^{\nu\sigma} + \bar g^{\mu\sigma} \bar g^{\nu\rho} \right) - \frac{1}{2} \bar g^{\mu\nu} \bar g^{\rho\sigma} \right] \, r_k(p^2/k^2) \, , \\
    \regulatoroperator_k^{A,\,\mu\nu}(p^2) &= p^2 \bar g^{\mu\nu} \, r_k(p^2/k^2) \, , \\
    \regulatoroperator_k^{\phi}(p^2) &= p^2 \, r_k(p^2/k^2) \, . \\
\end{aligned}
\end{equation}
In this work, we pick the shape function \cite{Litim:2001up}
\begin{align}
    r_k(y)=\left(\frac{1}{y}-1\right)\theta(1-y) \, .
\end{align}
This allows us to perform all loop integrals explicitly, and results in analytic flow equations for all couplings.

\section{Results: Vanishing ULDM-photon coupling in asymptotic safety and beyond}\label{sec:results}

In this section, we investigate whether the model in Eq.~\eqref{eq:matter_ansatz} has a \ac{UV} completion that is generated by quantum gravity fluctuations. We then study what the resulting predictions for the value of $\thcoupling$ are.

\subsection{Beta functions for the matter couplings}

We will analyze the system in the gauge choice $\beta=0$. We have checked that the qualitative picture that emerges below persists for all other admissible choices of gauge parameter $\beta$. This indicates the robustness of our results. The beta functions and anomalous dimensions read
\begin{subequations}
\begin{align}
\begin{split}
\beta_{\scalarmass^2} &= (-2 + \eta_\phi) \scalarmass^2 
                    + \frac{5 (6 - \eta_h) \scalarmass^2 g}{12\pi (1 - 2\lambda)^2} 
                    + \frac{(6 - \eta_h) \scalarmass^2 g}{18\pi (1 - \frac{4}{3}\lambda)^2} \\
                    &\qquad - \frac{2 (6 - \eta_h) \scalarmass^4 g}{9\pi (1 + \scalarmass^2)(1 - \frac{4}{3}\lambda)^2} - \frac{2 (6 - \eta_\phi) \scalarmass^4 g}{9\pi (1 + \scalarmass^2)^2 (1 - \frac{4}{3}\lambda)} +\frac{3(10-\eta_A)\thcoupling^2}{320\pi^2} \, , \label{eq:betamass}
\end{split}\\
\begin{split}
\beta_{\thcoupling} &= (1 + \eta_A + \frac{1}{2}\eta_\phi)\, \thcoupling
                    -\frac{\thcoupling (10-{\eta_A}) g}{6 \pi  (1-2 \lambda)}
                    +\frac{5 \thcoupling (8-{\eta_A}) g}{18 \pi  (1-2\lambda)}
                    -\frac{\thcoupling^3 (10-{\eta_A})}{320\pi ^2 (1+\scalarmass^2)} \\
                    &\qquad -\frac{\thcoupling(10-\eta_h) g}{12 \pi  (1-2 \lambda )^2}
                    +\frac{5\thcoupling (8-\eta_h) g}{18 \pi  (1-2 \lambda)^2}
                    -\frac{5\thcoupling (6-\eta_h) g}{18 \pi  (1-2\lambda )^2}
                    -\frac{\thcoupling^3 (10-\eta_{\phi})}{640\pi ^2 (1+\scalarmass^2)^2} \, , \label{eq:betaTh}
\end{split}\\
\begin{split}
\eta_\phi &= \frac{(4+ \eta_A)\, \thcoupling^2}{128\pi^2}
                    + \frac{(8 - \eta_h)\, g}{144\pi(1 + \scalarmass^2)(1 - \frac{4}{3}\lambda)^2}
                    + \frac{2 (6 - \eta_h)\, \scalarmass^2 g}{9\pi (1 + \scalarmass^2)(1 - \frac{4}{3}\lambda)^2} \\
                    &\qquad + \frac{(8 - \eta_\phi)\, g}{144\pi (1 + \scalarmass^2)^2 (1 - \frac{4}{3}\lambda)}
                    - \frac{4 \scalarmass^4 g}{3\pi (1 + \scalarmass^2)^2 (1 - \frac{4}{3}\lambda)^2} \, , \label{eq:etaphi}
\end{split}\\
\begin{split}
\eta_A &= -\frac{(8 - \eta_A)\, \thcoupling^2}{384\pi^2 (1 + \scalarmass^2)}
                    - \frac{(8 - \eta_\phi)\, \thcoupling^2}{384\pi^2 (1 + \scalarmass^2)^2} + \frac{5 (6 - \eta_h)\, g}{18\pi (1 - 2\lambda)^2} \\
                    &\qquad - \frac{5 (8 - \eta_h)\, g}{36\pi (1 - 2\lambda)^2}
                    - \frac{5 (8 - \eta_A)\, g}{36\pi (1 - 2\lambda)} \, . \label{eq:etaA}
\end{split}
\end{align}
\label{eq:coupled_flows}
\end{subequations}
We first study the dependence of the \thcoupling{} and \scalarmass{} critical exponents at the free fixed point $\thcouplingFP=\scalarmassFP=0$, as a function of the gravitational couplings. We then compare this to other works studying the relevance of dimension-five matter operators under the influence of gravity \cite{deBrito:2021akp, deBrito:2025ges}. We leave the study of a more realistic scenario that includes a state-of-the-art computation of the flow of the gravitational couplings for future work. Correspondingly, we will also set $\eta_h=0$.

\subsection{Consequences of asymptotic safety}\label{sec:relevance_plots}

\subsubsection{No ULDM-gauge interactions in asymptotic safety}

The coupling $\thcoupling$ breaks the global shift symmetry $\phi \rightarrow \phi + a$, as well as the global $\mathbb{Z}_2$-symmetry $\phi \rightarrow -\phi$, both of which are symmetries of the kinetic term, and thus of the minimal coupling between the scalar and gravity. As a consequence of the discussion in \cref{sec:global_symmetries}, we therefore expect that, if $\thcoupling$ is set to zero, it is not generated by quantum-gravity fluctuations. Our calculation explicitly confirms this, cf.~Eq.~\eqref{eq:betaTh}, because all terms in $\beta_{\thcoupling}$ are proportional to $\thcoupling$. This constitutes yet another example of how near-perturbative fluctuations in Euclidean asymptotic safety respect global symmetries.

Accordingly, the existence of a fixed point at $\thcouplingFP=0$ and $\scalarmassFP^2=0$ is guaranteed.\footnote{The existence of other, genuinely non-perturbative fixed points is of course not excluded. There is, however, no indication for it within our calculation. In fact, the beta function for $\thcoupling$ features a zero at $\thcouplingFP\neq 0$ precisely when $\thcoupling$ becomes relevant at the free fixed point. Our study therefore can also be interpreted in a less conservative way than we do, namely as the search for a region in the $(g, \lambda)$-plane where an interacting fixed point may exist for $\thcoupling$. We rush to add that establishing such a fixed point requires significantly more extended truncations and is not equivalent to showing relevance of $\thcoupling$ at the Gaussian fixed point, because this is only a necessary but not a sufficient condition for a non-trivial zero of the beta function to robustly exist.} Given that this is the free fixed point, one might jump to the conclusion that $\thcoupling$ is irrelevant and $\scalarmass^2$ is relevant. However, the scaling exponents of the free fixed point are dressed by quantum-gravity fluctuations. Thus, the scaling exponents contain the canonical dimensions of the couplings, and -- potentially in competition with these -- contributions $\sim g$. The main aim of our work is to calculate these contributions and determine their sign. From a fixed-point value $\thcouplingFP=0$, a non-zero value at low scales can only be reached if $\thcoupling$ corresponds to a relevant perturbation of the fixed point. Thus, the corresponding critical exponent must have a positive sign to avoid the prediction that $\thcoupling=0$ at all scales.

The critical exponents at the free matter fixed point are
\begin{align}
    \theta_{\scalarmass^2} &= 2-\frac{g_\ast}{3\pi(1-\frac{4}{3}\lambda_\ast)^2}-\frac{5g_\ast}{2\pi(1-2\lambda_\ast)^2}+\frac{16 g_\ast (2 \lambda_\ast -3)}{3(1-\frac{4}{3} \lambda_\ast) (g_\ast+144\pi(1-\frac{4}{3}\lambda_\ast))} \, , \label{eq:crit_GFP_free_param_grav_mass} \\
\begin{split}
    \theta_\thcoupling &= -1-\frac{g_\ast}{6\pi(1-2\lambda_\ast)^2}+\frac{8 g_\ast (2 \lambda_\ast -3)}{3(1-\frac{4}{3} \lambda_\ast) (g_\ast+144\pi(1-\frac{4}{3}\lambda_\ast))} \\
    &\quad+\frac{g_\ast (5 g_\ast+12\pi(10\lambda_\ast-4))}{3 \pi  (1-2 \lambda_\ast ) (5 g_\ast-36\pi(1-2\lambda_\ast))}\,.
\end{split}\label{eq:crit_GFP_free_param_grav}
\end{align}
The origin of the all-orders-in $g_{\ast}$ contributions lies in equations determining the anomalous dimensions, Eqs.~\eqref{eq:etaphi} and \eqref{eq:etaA}. When these are solved for $\eta_A$ and $\eta_{\phi}$, a rational dependence on $g_{\ast}$ appears.

To get some intuition for the expressions for the critical exponents, we first set $\lambda_{\ast}=0$ and expand to leading order in $g_{\ast}$, and obtain
\begin{align}
\theta_{\scalarmass^2}\Big|_{\lambda_{\ast}=0} &= 2- \frac{53}{18\pi}g_\ast+\mathcal{O}\left(g_{\ast}^2\right) \, , \\
\theta_{\thcoupling}\Big|_{\lambda_{\ast}=0} &= -1+\frac{2}{9\pi}g_{\ast} + \mathcal{O}\left(g_{\ast}^2\right) \, .
\end{align}
We recover the well-known result that the scaling exponent of the mass is shifted towards irrelevance, such that the mass changes its scaling behavior from relevant to irrelevant at a critical value of $g_{\ast}$, see, e.g., \cite{Narain:2009fy, Wetterich:2016uxm, Eichhorn:2020sbo}. By contrast, $\thcoupling$ is shifted towards relevance. However, the numerical prefactor of the contribution $\sim g_{\ast}$ is $\mathcal{O}(10^{-1})$, thus a relatively large value of $g_{\ast}$ is needed to achieve a change in sign in $\theta_{\thcoupling}$. While the value of $g_{\ast}$ by itself cannot be used to determine whether or not the theory is non-perturbative, the values of anomalous dimensions can be used to make that determination. In fact, $\eta_A=-\frac{5}{9\pi}g_{\ast}$ to leading order in $g_{\ast}$ at $\lambda_{\ast}=0$. Accordingly, we see that the values of $g_{\ast}$ needed to achieve a change in sign in $\theta_{\thcoupling}$ result in a large value of $\eta_A$, an indication of a truly non-perturbative regime, where our truncation -- based on the assumption of near-perturbativity -- is insufficient. We thus conclude that at $\lambda_{\ast}=0$, $\thcoupling$ is irrelevant at the fixed point at which $\thcouplingFP=0$. This inevitably leads to \emph{the prediction that $\thcoupling$ vanishes at all scales}.

\begin{figure}[t]
\centering
\includegraphics[width=0.7\textwidth]{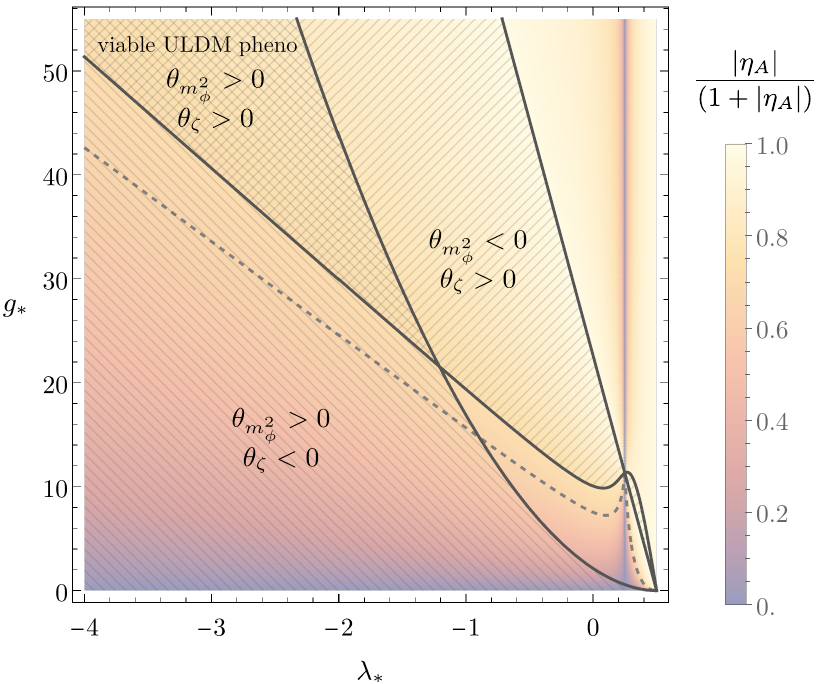} 
\caption{Regions of relevance of the critical exponents of the squared scalar mass and the \ac{ULDM}-photon-coupling, as a function of the gravitational sector fixed point values. Different types of hatchings indicate the regions where different critical exponents are relevant. In the unhatched region, both couplings are irrelevant. The gradient indicates the value of the photon anomalous dimension, with darker (lighter) colors indicating smaller (larger) values. The dashed line indicates where $|\eta_A|=2$, and delimits the region up until which we trust our approximation. In the diamond-hatched region, both couplings are relevant, and can thus yield viable \ac{ULDM} phenomenology. This region is however beyond the region of trust of the approximation.}
\label{fig:GFP_crits_lambda_g_LdW_gauge}
\end{figure}

Next, we consider the system away from $\lambda_{\ast}=0$ and to all orders in $g_{\ast}$. Our conclusions from the simpler analysis persist: large values of the gravitational coupling $g_\ast$ can induce a change of sign in the exponent $\theta_{\thcoupling}$. This is shown in \cref{fig:GFP_crits_lambda_g_LdW_gauge}, where we indicate the different regions where the two couplings are relevant by different choices of hatchings.\footnote{We note the change of sign of $\eta_A$ at $\lambda_\ast=\frac{1}{4}$, akin to the behavior described in \cite{Folkerts:2011jz, Christiansen:2017cxa} for non-Abelian gauge fields. This sign flip was shown to disappear when computing the momentum dependence of the gauge field anomalous dimension \cite{Christiansen:2017cxa}.} We find that $\theta_{\thcoupling}>0$ is only realized for large $g_\ast$, and in particular above the dashed line which indicates $|\eta_A|=2$. In this regime, $\eta_A$ grows large, such that, within the regime of validity of our truncation, we predict that $\thcoupling=0$ at all scales. At the same time, the mass remains relevant, such that the scalar field can become massive.

As a consequence, we find that \ac{ULDM}-photon couplings are not compatible with asymptotic safety, at least within the assumptions and approximations underlying our study. Because gravity is ``blind'' to internal symmetries, the gravitational contribution to the \ac{ULDM}-gluon coupling is the same as that for the photon. Accordingly, also that coupling vanishes at all scales. Thus, in an asymptotically safe theory, and in the absence of further \ac{BSM} sectors, which on their own generate such couplings, we predict that nuclear transitions are not sensitive to \ac{ULDM} through dimension-five couplings.

\subsubsection{Comparison of different dimension-five operators}

Several dimension-five operators play important roles in \ac{BSM} phenomenology, including, e.g., the \ac{ULDM}-photon coupling $\thcoupling$, the \ac{ALP}-photon coupling, as well as the Weinberg operator in the neutrino sector. These three operators have been studied in asymptotic safety~\cite{deBrito:2021akp, deBrito:2025ges}, and a comparison of their scaling exponents is informative.\footnote{These works investigated a similar setup as we did, namely treating the gravitational couplings as free parameters.} In particular, we aim at understanding whether the gravitational contribution to the scaling dimension of these interactions is similar across the three cases, such that we may form a hypothesis guiding future studies of further dimension-five operators. In fact, none of the three interactions is relevant and can thus be non-zero at low energies in a near-perturbative fixed-point regime, although all three are shifted towards relevance. For the analysis of the Weinberg operator in \cite{deBrito:2025ges}, the shift towards relevance holds in large parts of the parameter space, but is quantitatively very small. This is because the anomalous dimensions of scalars and fermions shift the interaction towards relevance, but the direct gravitational contribution to the vertex is towards irrelevance. Together, the contributions nearly cancel out.

\begin{figure}[t]
\centering
\includegraphics[width=0.7\textwidth]{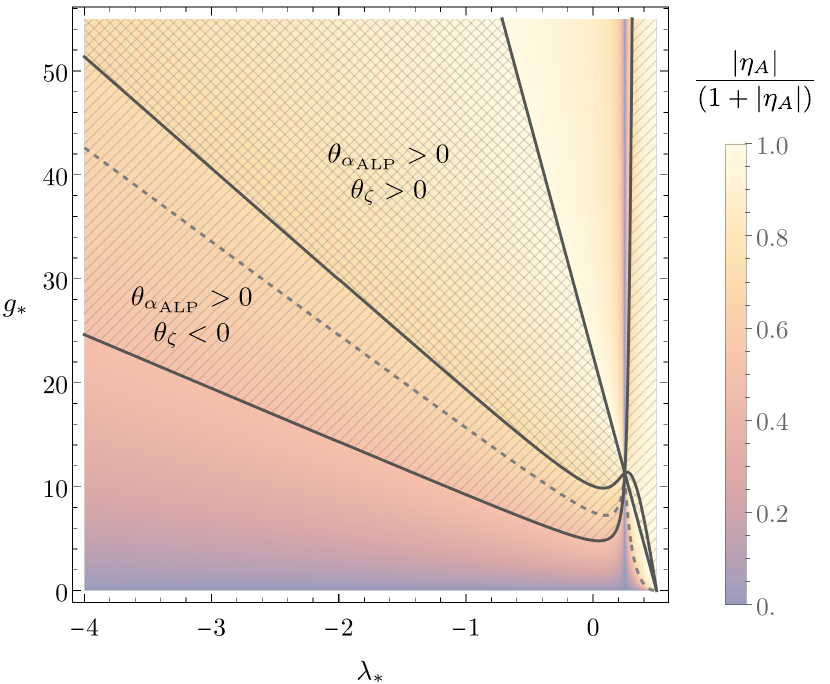} 
\caption{Regions of relevance of the \ac{ALP} and $\thcoupling$ coupling, as a function of the gravitational sector fixed point values. As in \cref{fig:GFP_crits_lambda_g_LdW_gauge}, the gradient indicates the value of the photon anomalous dimension, with darker (lighter) colors indicating smaller (larger) values. The dashed line indicates where $|\eta_A|=2$, and delimits the region up until which we trust our approximation.}
\label{fig:Th_ALP_Dim5_GFP}
\end{figure}

For the \ac{ULDM}-photon coupling and the \ac{ALP}-photon coupling, the shift towards relevance is again due to an anomalous dimension, in this case that of the gauge field. Accordingly, the \ac{ALP}-photon coupling, which has no direct gravitational contribution, turns relevant at the smallest value of $g_{\ast}$ of all three dimension-five operators, cf. \cref{fig:Th_ALP_Dim5_GFP}.

At a quantitative level, the significant difference between the lower boundaries\footnote{The upper bound of the relevance region in \cref{fig:Th_ALP_Dim5_GFP} is shared between the \ac{ALP} and $\thcoupling$ operator, since it corresponds to a singularity in the coupling flow located at $\lambda_\ast\leq0$ and $g^\text{boundary}_{\ast}=\frac{36\pi}{5}\left( 1-2\lambda_\ast\right)$, which drives the critical exponent to cross through infinity and change sign. This singularity is visible in Eq.~\eqref{eq:crit_GFP_free_param_grav}, and originates from contributions to the gauge anomalous dimension. However, this boundary lies deep outside of the near-perturbative region defined by $-2\leq\eta_A\leq2$, where we trust our approximation. Hence, we interpret it as a truncation artifact rather than a physical singularity.} of the regions of relevance of $\theta_{\thcoupling}$ and $\theta_{\alpha_\text{{ALP}}}$ are traced back to the topological nature of the dimension-five \ac{ALP} term,
\begin{equation}
    \Gamma_{k}^{\text{ALP}} \supset \int \text{d}^4x \,\alpha_\text{{ALP}}\, \epsilon^{\mu\nu\rho\sigma} \, \phi F_{\mu\nu}F_{\rho\sigma} \, ,
\end{equation}
where $\epsilon^{\mu\nu\rho\sigma}$ is the antisymmetric Levi-Civita tensor density, which makes the term topological. Diagrammatically, this means that no graviton-scalar-gauge vertex can be built \cite{deBrito:2021akp}, so the running of the corresponding coupling is extracted from the pure $\phi A A$-vertex. This leads to the simple flow 
\begin{equation}\label{eq:ALP_flow}
    \beta_{\alpha_\text{{ALP}}} = \left(1 + \eta_A + \frac{1}{2}\eta_\phi\right)\, \alpha_\text{{ALP}}\,.
\end{equation}
The additional terms appearing in Eq.~\eqref{eq:betaTh} compared to Eq.~\eqref{eq:ALP_flow} lead to a difference in critical exponents
\begin{equation}\label{eq:ThDM_vs_ALP_crit}
    \theta_{\thcoupling} - \theta_{\alpha_\text{{ALP}}} = -\frac{g_\ast}{6 \pi}\frac{1}{(1-2 \lambda_\ast)^2} - \frac{g_\ast}{3 \pi}\frac{12\pi(10\lambda_\ast-1)-5 g_\ast}{(1-2\lambda_\ast)(5 g_\ast-36 \pi(1-2 \lambda_\ast))} \, ,
\end{equation}
while the anomalous dimensions $\eta_{A,\phi}$ agree at the free matter fixed point between the pseudo-scalar and scalar \ac{ULDM} case. For all $g_\ast >0$ and $\lambda_\ast<0$, the additional terms in Eq.~\eqref{eq:ThDM_vs_ALP_crit} push the critical exponent $\theta_{\thcoupling}$ towards \emph{irrelevance}.

We extract the more general lesson, that the anomalous dimensions of the fields shift interactions towards relevance in large parts of the parameter space and large regions of choices of gauge parameters. Direct gravitational contributions to the scaling dimension may partially cancel this shift, as it happens, e.g., for the \ac{ULDM}-photon coupling and even more extremely for the Weinberg operator. However, from this analysis, it also becomes clear that if an interaction exists for which the direct contribution also results in a shift towards relevance, then a dimension-five operator may become relevant for significantly lower $g_{\ast}(\lambda_{\ast})$ than the boundary for the \ac{ULDM}-photon or the \ac{ALP}-photon coupling. We speculate that non-minimal dimension-five interactions of fermions and/or scalars may be interesting candidates to consider here.

\subsection{ULDM-gauge couplings in perturbative quantum gravity}

The implications of our results are not limited to asymptotic safety. Rather, our calculations also apply in the context of a perturbative \ac{EFT} for gravity, which is \ac{UV}-completed, e.g., in string theory, or some other setting. Then, in order to connect to General Relativity in the \ac{IR}, it is generically expected that a regime exists in which quantum gravity can be described by perturbative quantum fluctuations of the metric. In this regime, our beta functions apply, and $g$ and $\lambda$ should in that regime be understood as scale-dependent, and not restricted to take fixed-point values.

Within such a regime, we are not interested in fixed points in $\thcoupling$, but rather care about two questions: First, if $\thcoupling$ is zero at the onset of such a regime, is it still zero at its end? In other words, does perturbative quantum gravity induce $\thcoupling$? Second, if $\thcoupling$ has some size $\thcoupling/\Lambda_\text{UV}$, where $\Lambda_\text{UV}$ is the scale at which the perturbative regime starts, then what is its size at the end of this regime? In other words, do perturbative quantum gravity fluctuations increase or decrease the coupling, if it is already nonzero due to the properties of the underlying \ac{UV}-complete theory?

We find that the answer to the first question is negative, i.e., $\thcoupling$ cannot be generated by perturbative quantum-gravity fluctuations. Thus, if it is nonzero due to quantum gravity, it must be induced in some (potentially non-perturbative) regime, in which quantum gravity is described by a string theory, a theory of discrete spacetime, or some other non-quantum-field-theoretic \ac{UV} completion.

Regarding the second question, we consider the \ac{RG} flow in the perturbative regime in the presence of the anomalous scaling dimension generated by quantum gravity fluctuations. To simplify our argument, we will first assume that the scaling dimension is constant and later explain what changes if the scaling dimension is itself scale-dependent and therefore not an actual scaling dimension. We have that
\begin{equation}
\frac{\thcoupling(k)}{k} = \frac{\thcoupling(\Lambda_\text{UV})}{\Lambda_\text{UV}}\left(\frac{k}{\Lambda_\text{UV}}\right)^{d_{\thcoupling}} \, ,
\end{equation}
where 
\begin{equation}\label{eq:pertscaling}
d_{\thcoupling} = \frac{\partial \beta_{\thcoupling}}{\partial \thcoupling}-1 \, ,
\end{equation}
i.e., $d_{\thcoupling}$ would be the anomalous scaling dimension in a regime in which all couplings were constant, that is, at a fixed point. From this expression, we see that the value of $\thcoupling$ at the end of the perturbative quantum-gravity regime, where quantum-gravity fluctuations become negligible, is \emph{increased} by the factor $\left(\frac{k}{\Lambda_\text{UV}}\right)^{d_{\thcoupling}}>1$, if $d_{\thcoupling}$ is negative.

If the gravitational couplings are scale-dependent, as they generically are in the perturbative regime, then the simple power-law \eqref{eq:pertscaling} no longer holds and the enhancement or suppression factor has to be calculated by actually integrating \ac{RG} trajectories. However, we see that the sign of $d_{\thcoupling}$ determines whether or not the coupling will be enhanced.

In our asymptotically safe computation, we generically find a negative sign, so the coupling is generically enhanced in the perturbative quantum gravity regime, as long as it is nonzero at the onset of this regime. From this, we tentatively conclude that perturbative quantum gravity fluctuations make it simpler to detect $\thcoupling$ (by enhancing it), once it has been induced by a \ac{UV} completion of perturbative quantum gravity.

In the specific \ac{UV} completion constituted by asymptotic safety, it is not induced, and thus its enhancement plays no role. However, in other \ac{UV} completions the situation may be different.

\section{Discussion and outlook}\label{sec:conclusions}

In this work, we considered the impact of asymptotically safe quantum gravity on the running of the dimension-five \ac{ULDM} scalar-photon interaction. As a dimension-five interaction, the associated coupling is dimensionful, and may be written as a dimensionless coefficient divided by a scale, $d/\Lambda_\text{BSM}$, commonly interpreted as the mass scale of the \ac{BSM} physics that induces this interaction. This interaction may be induced by new physics below the Planck scale, or by gravitational physics in the trans-Planckian regime. The former possibility will be tightly constrained by nuclear-clock experiments using a transition in Thorium, which may even in the future constrain the \emph{dimensionful} coupling $d/\Lambda_\text{BSM}$ to scales above the Planck scale. Thus, to satisfy these projected constraints, new physics at a mass scale below the Planck scale would have to produce a highly suppressed dimensionless coefficient $d \ll 1$. Following naturalness arguments, the alternative possibility that the $\thcoupling$ operator is only induced by quantum gravity beyond the Planck scale is preferred (see, however, \cite{Hook:2018jle, Brzeminski:2020uhm}). 

This is the possibility we tentatively rule out within a specific quantum-gravity theory, by predicting that this coupling vanishes exactly in asymptotically safe quantum gravity.

Crucially, this is a testable prediction, leveraging that Thorium-based optical nuclear clocks are projected to bound the $\thcoupling$ coupling beyond the Planck scale \cite{Fuchs:2024xvc}.

Our results hold under several assumptions, some of a technical and others of a physical nature. On the technical side, key assumptions are that our Euclidean calculation carries over to a Lorentzian regime and that our truncation suffices to capture the physics of a near-perturbative regime. On the physical side, we assume that asymptotic safety is indeed near-perturbative. We also assume that the \ac{ULDM}-photon-interaction (and, similarly, the \ac{ULDM}-gluon interaction) does not have an interacting fixed point without gravity which would be dressed by gravitational fluctuations. Such a fixed point would necessarily be non-perturbative and we consider such a possibility unlikely. Finally, we also assume that the dark sector is minimalistic and that there are in particular no additional fields in the dark sector which could contribute to the \ac{ULDM}-photon-interaction in such a way as to make it relevant or shift its fixed-point value (and consequently also its low-energy value) away from zero.

Besides our prediction of a vanishing \ac{ULDM}-photon and \ac{ULDM}-gluon coupling, we also extract more structural lessons about asymptotic safety from our study. To the existing studies of dimension-five operators, we have added another example and conclude that there is a very slim chance that a dimension-five operator becomes relevant in asymptotic safety, if the signs of all contributions (anomalous dimensions of the fields and anomalous scaling contribution to the interaction) are the same and are all in the direction of relevance. A more likely scenario, however, is that dimension-five operators are irrelevant in asymptotic safety, further consolidating the predictive power of this scenario for particle physics.

Further, we draw conclusions for the perturbative regime of quantum gravity, where it constitutes an \ac{EFT} and may be \ac{UV}-completed by, e.g., string theory, or some other non-quantum-field-theoretic theory. In this regime, we find that perturbative quantum gravity does not induce $\thcoupling$, but \emph{enhances} it -- thus making a detection more likely -- if it is induced within the \ac{UV}-complete quantum-gravity theory.

In the future, our study could be extended by explicitly computing the flow of gravitational couplings, and checking where the fixed point lies in \cref{fig:GFP_crits_lambda_g_LdW_gauge}. By considering the impact of results on anomalous dimensions, including gravitational ones, we could also verify the near-perturbative assumption we made to bootstrap the approximated solution to the flow equation, Eq.~\eqref{eq:wetterich_eq}. Finally, a computation in Lorentzian signature, albeit challenging \cite{DAngelo:2023tis,Bonanno:2021squ, Fehre:2021eob, Saueressig:2023tfy, DAngelo:2023wje, Ferrero:2024rvi, Saueressig:2025ypi, DAngelo:2025yoy, Pawlowski:2025etp,Banerjee:2022xvi, Banerjee:2024tap, Thiemann:2024vjx}, is required to more robustly make contact with experimental bounds. This would both be one of the first Lorentzian gravity-matter studies in asymptotic safety \cite{Kher:2025rve}, and a useful comparison to our Euclidean prediction for $\thcoupling$.

\paragraph{Acknowledgements}

We acknowledge a helpful exchange with M.~Schiffer and M.~Fil\-zing\-er, and Laura Blackburn. G.~A.~was supported by the Science Technology and Facilities Council (STFC) under the Studentship Grant ST/X508822/1. G.~A.~is grateful for the hospitality of Heidelberg University during the early stages of this project.
A.~E.~acknowledges the European Research Council's (ERC) support under the European Union’s Horizon 2020 research and innovation program Grant agreement No.~101170215 (ProbeQG). A.~E.~is also supported by the Deutsche Forschungsgemeinschaft (DFG, German Research Foundation) under Germany’s Excel- lence Strategy EXC 2181/1 - 390900948 (the Heidelberg STRUCTURES Excellence Cluster).

\bibliographystyle{JHEP}
\bibliography{bibliography.bib}

\end{document}